# A Metastable Pentagonal 2D Material Synthesized by Symmetry-Driven Epitaxy

Lina Liu[1,4,7],[#] Yujin Ji[2],[#] Marco Bianchi[3], Saban M. Hus[5], Zheshen Li[6], Richard Balog[7], Jill A. Miwa[3], Philip Hofmann[3], An-ping Li[5], Dmitry Y. Zemlyanov[4],[*] Youyong Li[2],[*] Yong P. Chen[1,4,7,8*]

[1]Department of Physics and Astronomy, Purdue University, West Lafayette, Indiana, 47907, USA

[2]Institute of Functional Nano and Soft Materials, Soochow University, Suzhou, Jiangsu Province, 215123, China

[3]Department of Physics and Astronomy and Villum Center for Dirac Materials, Aarhus University, 8000, Aarhus C, Denmark

[4]Birck Nanotechnology Center, Purdue University, West Lafayette, Indiana, 47907, USA

[5]Center for Nanophase Materials Sciences, Oak Ridge National Laboratory, Oak Ridge, TN, 37831, USA

[6]Department of Physics and Astronomy and center for Storage Ring Facilities (ISA), Aarhus University, 8000, Aarhus C, Denmark

[7]Department of Physics and Astronomy and Villum Center for Hybrid Quantum Materials and Devices, Aarhus University, 8000, Aarhus C, Denmark

[8]Purdue Quantum Science and Engineering Institute and School of Electrical and Computer Engineering, Purdue University, West Lafayette, Indiana, 47907, USA

Corresponding to: yongchen@purdue.edu

yyli@suda.edu.cn

dzemlian@purdue.edu

## Abstract

Most two-dimensional (2D) materials experimentally studied so far have hexagons as their building blocks. Only a few exceptions, such as $PdSe_2$, are lower in energy in pentagonal phases and exhibit pentagons as building blocks. While theory has predicted a large number of pentagonal 2D materials, many of them are metastable and their experimental realization is difficult. Here we report the successful synthesis of a metastable pentagonal 2D material, the monolayer pentagonal $PdTe_2$, by symmetry-driven epitaxy. Scanning tunneling microscopy and complementary spectroscopy measurements are used to characterize the monolayer pentagonal $PdTe_2$, which demonstrates well-ordered low-symmetry atomic arrangements and is stabilized by lattice matching with the underlying Pd(100) substrate. Theoretical calculations, along with angle-resolved photoemission spectroscopy, reveal monolayer pentagonal $PdTe_2$ is a semiconductor with an indirect bandgap of 1.05 eV. Our work opens an avenue for the synthesis of pentagon-based 2D materials and gives opportunities to explore their applications such as multifunctional nanoelectronics.

Since the first few pentagonal 2D materials, pentagonal graphene[1] and noble metal dichalcogenides[2], were predicted several years ago, research interest has been growing in this subgroup of 2D materials due to their unique properties and potential applications in nanoelectronics, optoelectronics and thermoelectrics[3,4]. Unlike the much more studied hexagon-based 2D materials, the building blocks for pentagonal 2D materials are pentagons. Due to the well-known pentagonal tiling rule, where regular pentagons cannot tile a plane, pentagonal 2D materials tend to form puckered layers. Such a puckered pentagon-based structure presents low crystallographic symmetry, leading to an orthorhombic crystal structure with rectangular unit cells[5]. The low crystallographic symmetry introduces large in-plane anisotropy and low thermal conductivity, making them promising candidate materials for future anisotropic electronics and thermoelectrics[6-10]. The puckered layers in pentagonal 2D materials also induce renormalization of bond lengths and angles, resulting in decreased layer thickness and increased layer flexibility[11]. This makes pentagonal transition metal dichalcogenides (TMDCs) monolayers ~40-50% thinner compared to hexagonal TMDC monolayers[11] and could facilitate applications in flexible devices and wearable electronics. Furthermore, many other fascinating properties have been predicted for pentagonal 2D materials such as ultrahigh strength of pentagonal graphene[1], room temperature quantum spin Hall effect of pentagonal $SnX_2$ (X=S, Se, Te)[12], magnetic Dirac fermions of pentagonal $MoS_2$[13] and ferromagnetism of semiconducting pentagonal $VTe_2$[14], indicating that the pentagonal 2D materials are not only promising materials for novel applications but also demonstrate great potentials in fundamental studies of new quantum states and phenomena.

Although pentagonal 2D materials have been theoretically predicted for years, few of them have been experimentally studied so far[3]. Only recently, $PdSe_2$ together with analogous PdPSe and PdPS, a few materials with pentagonal structures as thermodynamically stable phases, start to be investigated[15-18] (note their bulk crystals, with the orthorhombic structure, were studied much earlier[19,20]). Their properties including high mobility and large in-plane anisotropy have been unveiled, which facilitates their applications in functional nanoelectronics[15-18]. In addition, giant non-linear optical activity has been observed in $PdSe_2$[21,22], implying its potential in optoelectronics. More interestingly, the pentagonal structure offers $PdSe_2$ great flexibilities for structural reconstruction, which provides opportunities for phase engineering[23] and enables interlayer manipulation[24]. However, in contrast to $PdSe_2$, many pentagon-based 2D materials are metastable[25] (Supplementary Fig. 1), which makes the direct synthesis of these materials more challenging. Although previously some metastable phases of hexagon-based 2D TMDCs (such as 1T and 1T′ phases of $MoS_2$) have been synthesized[26-28], the direct synthesis of pentagon-based metastable phases has never been demonstrated. The lack of proper ways to synthesize and stabilize pentagonal 2D materials (particularly the metastable ones) has largely hindered the exploration of this unique class of 2D materials.

Here, we report the direct synthesis of metastable monolayer pentagonal $PdTe_2$. So far, all synthesized $PdTe_2$ (both bulk and 2D layers) demonstrate hexagonal 1T structures[29-31] since its hexagonal phase is energetically favorable with the lowest formation energy[32]. Based on our density functional theory (DFT) calculations, the formation energy for monolayer pentagonal $PdTe_2$ is only 0.04 eV/atom higher than that of the monolayer hexagonal $PdTe_2$ (Supplementary Fig. 1), suggesting potential feasibility to synthesize the pentagonal phase. Here, we use epitaxial growth, which is an efficient way to grow high-quality 2D materials[33,34], for the synthesis of pentagonal $PdTe_2$. During epitaxy, the substrate symmetry plays a crucial role and the as-grown structure can be well-controlled and stabilized due to lattice match with the substrate[35]. We demonstrated the synthesis of monolayer pentagonal $PdTe_2$ by symmetry-driven epitaxy through direct tellurization of Pd(100) surface, whose lattice exhibits good lattice-match with that of pentagonal $PdTe_2$. The successful growth of monolayer pentagonal $PdTe_2$ was confirmed by various structural and spectroscopic characterizations. Comprehensive scanning tunneling microscopy (STM) measurements and simulations, along with low electron energy diffraction (LEED) proved the atomic structures of pentagonal $PdTe_2$. X-ray photoelectron spectroscopy (XPS) measurements verified the formation of $PdTe_2$ and the monolayer thickness. Phonon dispersions were revealed by DFT calculations and corresponding lattice vibrational modes were observed in the high-resolution electron energy loss spectroscopy (HREELS) spectra. The DFT calculations revealed that the monolayer pentagonal $PdTe_2$ is a semiconductor with an indirect bandgap of 1.05 eV, which is consistent with the scanning tunneling spectroscopy (STS) result. Meanwhile, the valence bands of the monolayer pentagonal $PdTe_2$ were measured by angle-resolved photoemission spectroscopy (ARPES). The direct synthesis, along with the comprehensive measurements of atomic structures, phonon dispersions and electronic structures of monolayer pentagonal $PdTe_2$, will greatly accelerate the research field of pentagonal 2D materials.

We selected different facets of Pd single crystals as substrates for the epitaxial growth of different phases of $PdTe_2$. By taking advantage of the symmetries of the underlying substrates, hexagonal and pentagonal monolayer $PdTe_2$ were successfully synthesized through direct tellurization of Pd(111) and Pd(100) surfaces, respectively (Fig. 1a). The as-grown hexagonal monolayer $PdTe_2$ exhibits conventional 1T structure (Fig. 1b and 1c) with $(\sqrt{3}\times\sqrt{3})R30^{o}$ atomic arrangements, which has been thoroughly discussed in our previous work[36].

In this work, we focus on the pentagonal phase of $PdTe_2$ grown on Pd(100). Before growth, the surface of Pd(100) was freshly cleaned by repeated sputtering and annealing in ultrahigh vacuum (UHV). After cleaning, Pd atoms on the surface were full of dangling bonds and chemically reactive. Then Te was deposited on the Pd(100) substrate. Annealing at 500 $^{o}$C induced the formation of pentagonal $PdTe_2$, which as shown below exhibits lattice-match with the Pd(100) substrate (details of growth could be found in the Supplementary Fig. 2). This lattice-match serves as the driving force for the growth of the pentagonal phase of $PdTe_2$. Pentagonal $PdTe_2$ presents low crystallographic symmetry with monoclinic space group[8] of $P2_1/c$, which is significantly different from the hexagonal $PdTe_2$ with high symmetry space group[37] of $P\bar{3}m1$. Its out-of-plane geometries are puckered (Fig. 1a) and its projected in-plane atomic arrangements demonstrate pentagons as the building blocks (Fig. 1d). The projected in-plane structure exhibits a rectangular unit cell (Fig. 1d), with a tetra-coordination[5,15] for Pd atoms (where each Pd atom bonds with four Te atoms, two from top and bottom, respectively). For the out-of-plane structure, each top Te atom pairs with a bottom Te atom to form a Te-Te dimer across the layer[5], resulting in a puckered layer with a shorter vertical height of 1.69 Å compared with that (2.76 Å) of the monolayer hexagonal $PdTe_2$[11]. The atomic-resolution STM image taken after growth reveals one-dimensional zigzag-chains formed by bright dots (Fig. 1e), indicating low-symmetry arrangements of surface atoms. This structure agrees well with the patterns of topmost Te atoms of pentagonal $PdTe_2$. The measured unit cell of the as-grown structure is square (6.2×6.2 Å), which is slightly different from the previously calculated rectangular lattice for the free-standing monolayer pentagonal $PdTe_2$ with the unit cell of 6.1×6.4 Å[8,11]. The reason for this difference can be ascribed to the lattice confinement from the underlying Pd(100) substrate with a square lattice. We note that unlike the Pd(100) lattice, the as-grown $PdTe_2$ lattice is not invariant upon 90° rotation even when it has a square unit cell. According to the Pd-Te phase diagram[38] and theoretical database of 2D materials[39,40], for all known Pd-Te compounds, the pentagonal $PdTe_2$ we discuss here ($\beta$ pentagonal phase[3]) is the only one that has a rectangular-shape unit cell with lattice constants close to 6.2×6.2 Å, consistent with our STM images. While other types of pentagon-based phases have also been predicted for other materials (such as $\gamma$ phase for $PdS_2$)[3,41], the $\beta$ pentagonal phase we discuss here is the only one where zigzag chains could be formed by topmost atoms, again consistent with our STM observations thus confirming our structural assignment. We also notice that the as-grown monolayer pentagonal $PdTe_2$ is stable after air exposure (Supplementary Fig. 3), facilitating its potential applications in ambient conditions.

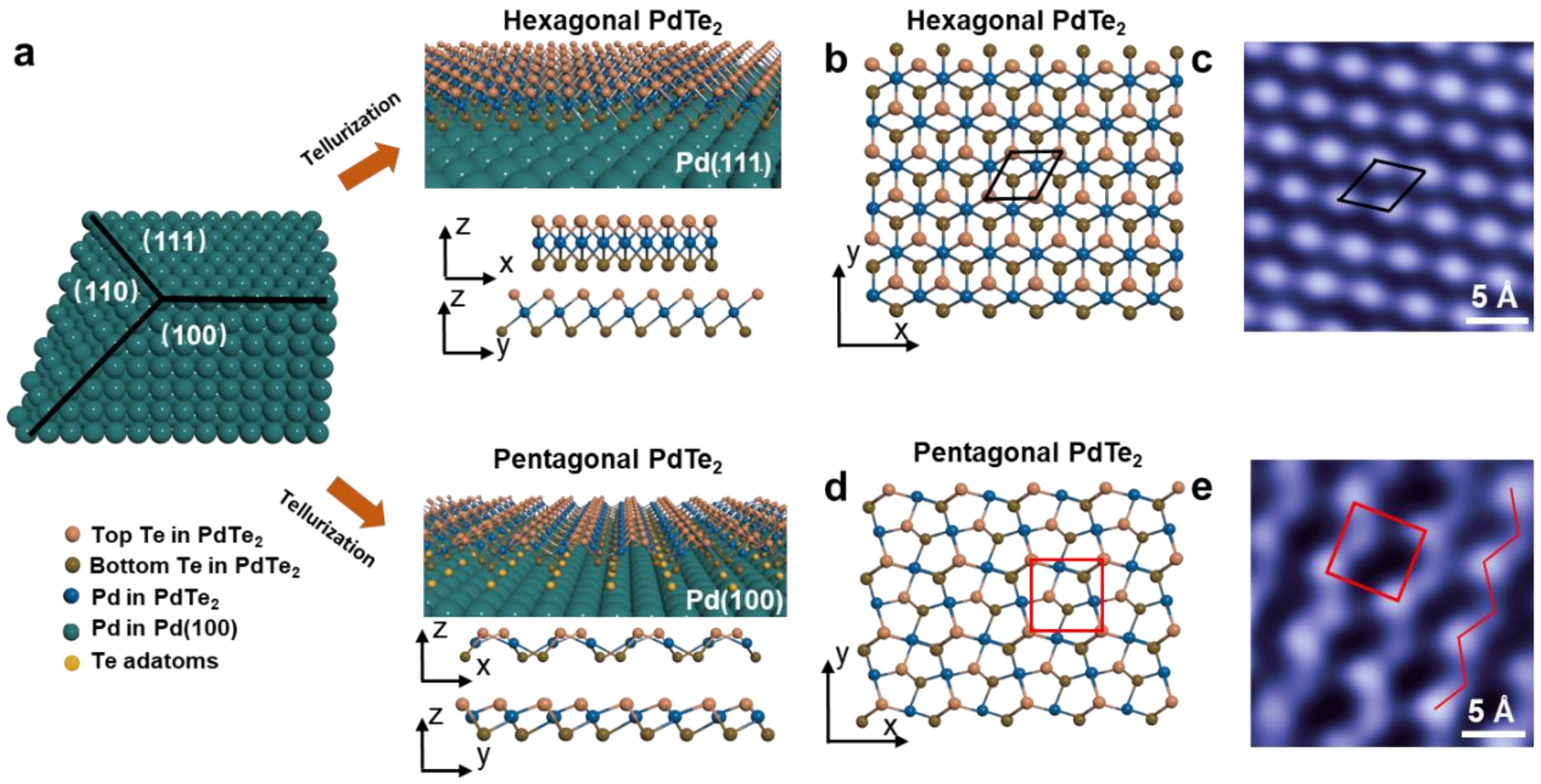

**Fig. 1:** Synthesis of monolayer hexagonal and pentagonal $PdTe_2$. **a**, Schematics of epitaxial growth of monolayer hexagonal and pentagonal $PdTe_2$ by direct tellurization of Pd(111) and Pd(100) surfaces, respectively. This schematic image is for the demonstration of key aspects of the growth method only. Te adatoms in the schematic image of pentagonal $PdTe_2$ on Pd(100) are discussed in the later text. **b** and **c**, A schematic lattice image (top view) and an atomic-resolved STM image of monolayer hexagonal $PdTe_2$, respectively. **d** and **e**, A schematic lattice image (top view) and an atomic-resolved STM image of monolayer pentagonal $PdTe_2$, respectively. Corresponding side views of the lattices shown in **b** and **d** are shown in **a**. Measurement parameters: sample bias voltage $V_b$ = 0.8 V, tunneling current magnitude $I_t$ = 0.7 nA for **c** and $V_b$ = -0.5 V, $I_t$ = 0.5 nA for **e**. Black rhombuses and red squares represent unit cells of hexagonal and pentagonal $PdTe_2$, respectively. Red lines in **e** indicate zigzag chains formed by topmost Te atoms.

We carried out synchrotron-based XPS measurements to verify the formation of $PdTe_2$. We observed two sets of peaks in the Pd 3*d* spectra: one set presents lower binding energies at 335.1 eV and 340.4 eV while the other presents higher binding energies at 336.0 eV and 341.3 eV (Supplementary Fig. 4a). The sets at lower and higher binding energies can be attributed to bulk Pd substrate and pentagonal $PdTe_2$, respectively. Two sets of Te 4*d* peaks were also observed in the XPS spectra (Supplementary Fig. 4b). The lower binding energies of 40.6 eV and 42.1 eV are ascribed to Te in $PdTe_2$. The existence of the set of higher binding energies of 41.5 eV and 43.0 eV indicates that there is another type of Te in the system. We attribute it to a Te adatoms layer between the as-grown pentagonal $PdTe_2$ and the Pd(100) substrate, supported by previous works in similar systems (e.g. TMDC monolayers epitaxially grown on corresponding single crystals)[42,43] and our DFT calculations (A comparison between the XPS spectra of as-grown pentagonal and hexagonal $PdTe_2$ demonstrates that Te adatoms are unique in pentagonal $PdTe_2$/Pd(100), Supplementary Fig. 4c. More discussions about the Te adatoms could be found in Supplementary Fig. 5, Supplementary Table 1 and Supplementary Fig 6). Moreover, we carried out the XPS measurements under both normal and grazing emission angle to determine the relative contribution of $PdTe_2$, Te adatoms and the Pd(100) substrate (Supplementary Fig. 4a and 4b). The results show the Pd peak area ratio of $PdTe_2$ to bulk Pd substrate became bigger at grazing emission while the Te peak area ratio of Te adatoms to $PdTe_2$ became smaller, suggesting that $PdTe_2$ is on top of Pd(100) and Te adatoms are under $PdTe_2$. Based on our XPS results, the thickness of the as-grown pentagonal $PdTe_2$ is estimated[36,44] to be ~1.6 Å, consistent with monolayer pentagonal $PdTe_2$[11]. This value is also much smaller than the calculated thickness of bilayer pentagonal $PdTe_2$ (5.1 Å), indicating the as-grown $PdTe_2$ is unlikely a bilayer.

In addition, we performed DFT calculations and HREELS measurements to explore the lattice dynamics of monolayer pentagonal $PdTe_2$. The calculated phonon dispersions of monolayer pentagonal $PdTe_2$ (Fig. 2a) show dramatically different features from those of monolayer hexagonal $PdTe_2$ (Supplementary Fig. 7a). In the meantime, HREELS was utilized to experimentally probe the phonon modes. Two components could be resolved at 23 meV and 26 meV in the energy loss spectrum measured at the Γ point (Fig. 2b), in fair agreement with the calculated optical phonon modes of monolayer pentagonal $PdTe_2$ and notably different from the measured HREELS spectrum (which shows peaks at 15 meV and 25 meV) in monolayer hexagonal $PdTe_2$ (Supplementary Fig. 7b). No modes were observed on the Pd(100) substrate in the same energy region (Supplementary Fig. 8), proving the observed energy loss modes are from the as-grown pentagonal $PdTe_2$.

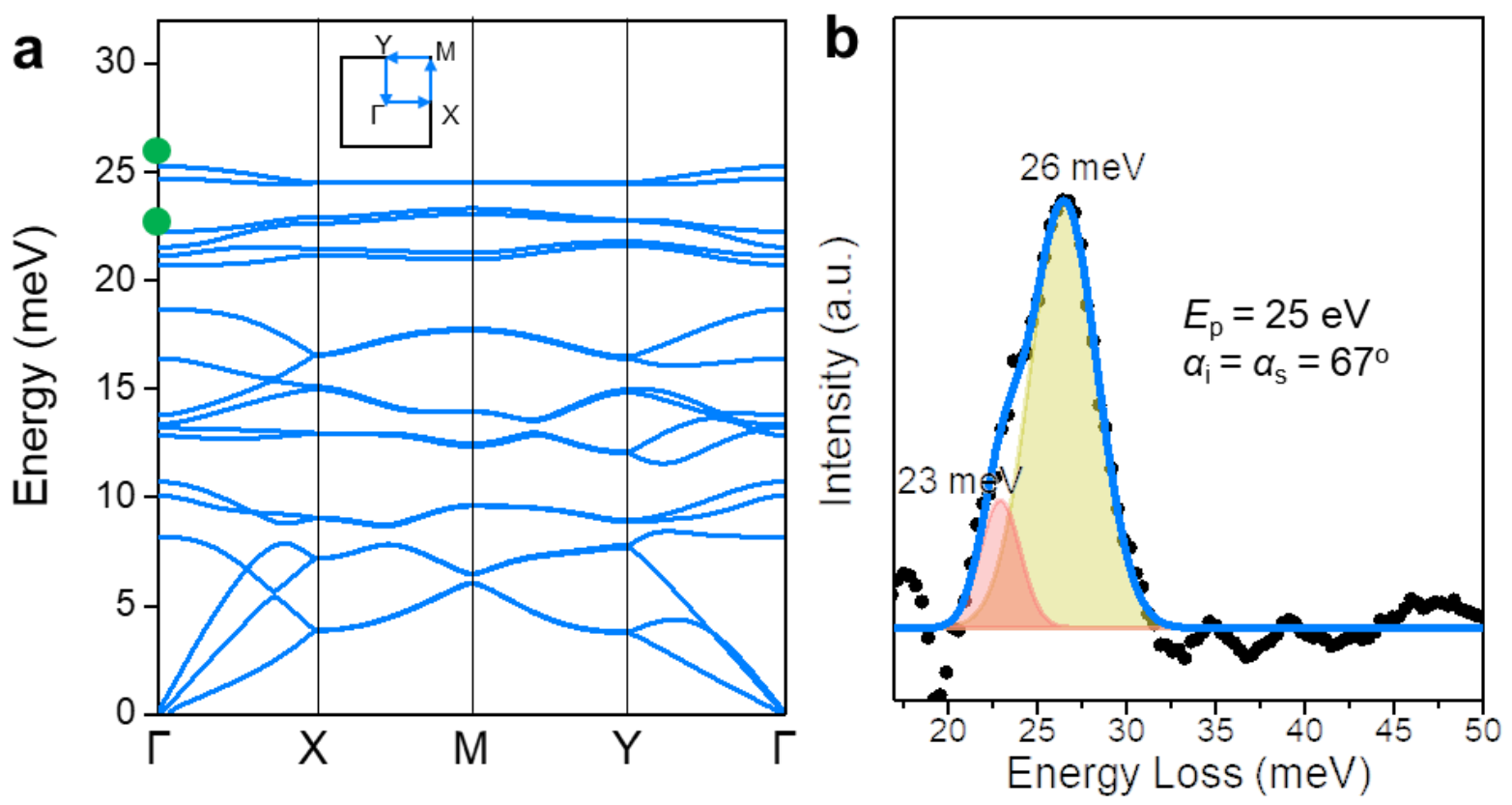

**Fig. 2:** Phonons of monolayer pentagonal $PdTe_2$. **a**, Calculated phonon dispersions of monolayer pentagonal $PdTe_2$ along the Γ-X-M-Y-Γ path of the Brillouin zone as shown in the inset. **b**, An electron energy loss spectrum of as-grown pentagonal $PdTe_2$ at the Γ point. Dashed black and solid blue curves are the original data curve (after background subtraction) and the fitting curve, respectively. Red and yellow peaks (at 23 meV and 26 meV respectively) are the fitting components and their energies are marked as green dots in **a**. The primary electron beam energy ($E_p$) is 25 eV. Incident and scattering angles ($\alpha_i$ and $\alpha_s$, respectively) are 67°.

For a comprehensive understanding of the atomic structures, the as-grown pentagonal $PdTe_2$ was studied by STM and LEED. After growth, the surface as seen by STM is clean and flat over a large area with sometimes an apparent step height of ~1.9 Å (Fig. 3a), consistent with a monoatomic step terrace of Pd(100) and indicating the monolayer $PdTe_2$ is epitaxially grown on the terrace. Note all the terraces we observed by STM exhibit a vertical height around this value, consistent with the imaged Pd(100) surface being fully covered by monolayer $PdTe_2$ with no bilayer or multilayer regions found. The magnified STM image shows clear and parallel zigzags chains, implying the well-ordered structures and high-quality of the as-grown pentagonal $PdTe_2$ (Fig. 3b). Individual atoms were clearly resolved in the zoomed-in STM image (Fig. 3c), confirming the low-symmetry atomic arrangements of the monolayer pentagonal $PdTe_2$. The unit cell of pentagonal $PdTe_2$ is square with dimensions of 6.2×6.2 Å, corresponding to the $(\sqrt{5}\times\sqrt{5})R26.6^{o}$ supercell of the underlying Pd(100) ($\sqrt{5}\times 2.75 = 6.15 \approx 6.2$ Å, where the lattice constant of the surface atoms of Pd(100) is 2.75 Å). Notably, even with a square unit cell, the as-grown pentagonal $PdTe_2$ exhibits only a two-fold (180$^{o}$) rotational symmetry rather than four-fold (90$^{o}$) rotational symmetry of the square lattice of the underlying Pd(100). Occasionally, coexistence of two differently-orientated domains in the as-grown pentagonal $PdTe_2$ could be observed (Fig. 3d). The angle between the two orientations is ~53$^{o}$ (approximately twice of 26.6$^{o}$), indicating that each orientation is rotated symmetrically by ~26.6$^{o}$ with respect to the lattice orientation of Pd(100). The *a* axes (the direction along zigzag chains) of the two orientations are along the $[01\overline{2}]$ and $[0\overline{1}\overline{2}]$ crystallographic directions of the Pd(100) topmost surface respectively, illustrating a good epitaxy (Fig. 3e). Due to the lack of four-fold (90$^{o}$) rotational symmetry, there could also exist two other possible orientations for the as-grown pentagonal $PdTe_2$ lattices by rotating the two observed orientations (Fig. 3d) by 90$^{o}$ (the two other possible lattice vectors are along the $[0\overline{2}\overline{1}]$and $[0\overline{2}1]$ directions as shown in Supplementary Fig. 9, though we have not yet observed $PdTe_2$ domains along these two directions, possibly due to the small scanning areas of STM). Furthermore, LEED characterizations were carried out to examine the crystallinity and orientation of the as-grown pentagonal $PdTe_2$ over a millimeter scale. As shown in Fig. 3f, bright and sharp diffraction spots were observed, proving the high crystallinity of the as-grown pentagonal $PdTe_2$ layer over a large surface area. Two sets of the diffraction patterns with a rotation angle of ~53$^{o}$ between them were clearly identified, which is consistent with the rotation angle between the two domains observed in STM (Fig. 3d). Comparing the LEED patterns of pentagonal $PdTe_2$ with those of bare Pd(100) (Supplementary Fig. 10), it is further confirmed that each set of the $PdTe_2$ diffraction patterns is rotated by ~26.6$^{o}$ with respect to the diffraction pattern (lattice orientation) of Pd(100). Note the LEED patterns obtained on the pentagonal $PdTe_2$ do possess a four-fold (90$^{o}$) rotational symmetry thus cannot distinguish two orientations that are rotated by 90$^{o}$ from each other (in other words the observed LEED patterns could be consistent with two to four different orientations coexisting on the surface measured).

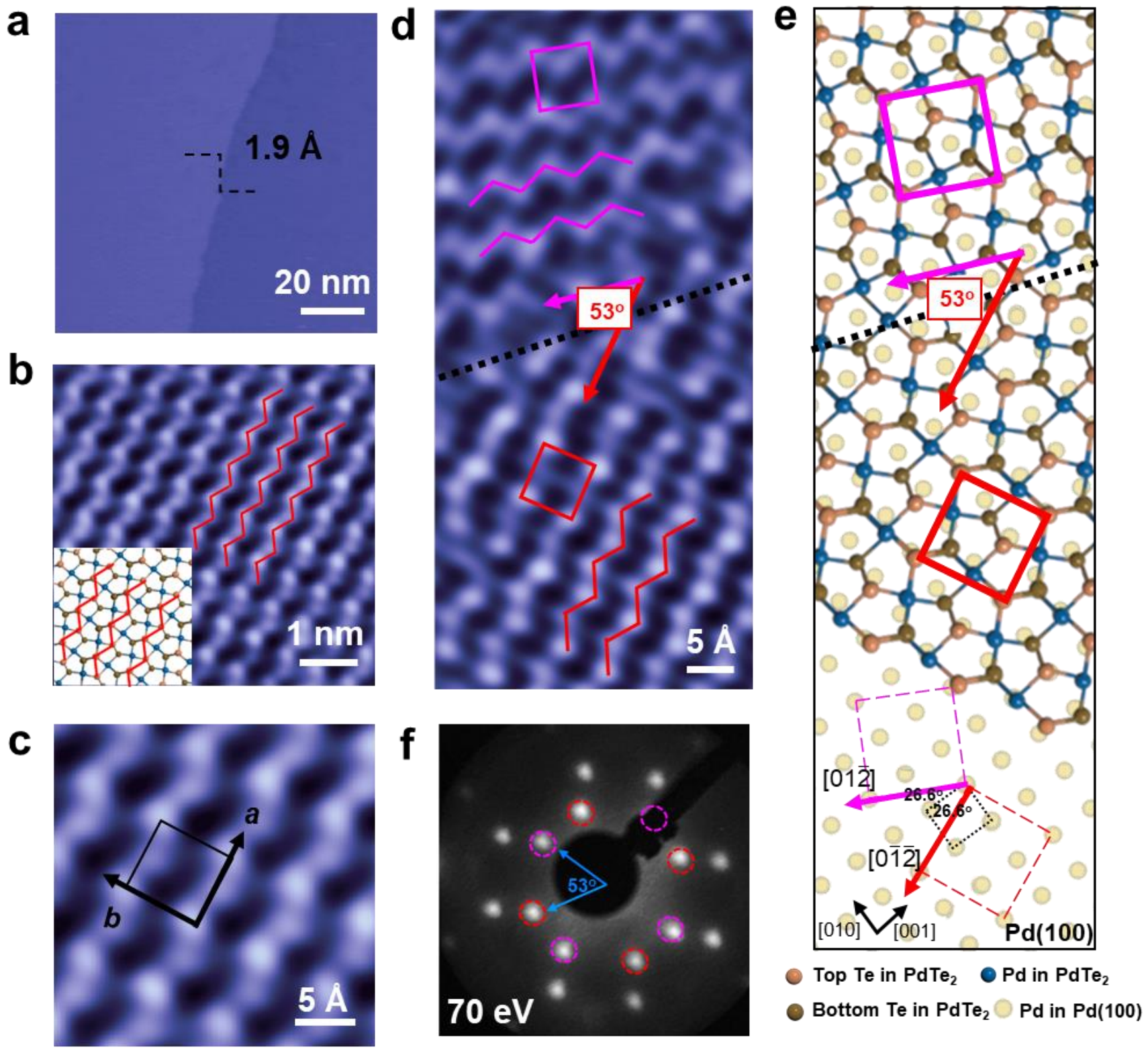

**Fig. 3:** STM and LEED measurements of the as-grown monolayer pentagonal $PdTe_2$. **a-c**, Topographic (**a**), magnified (**b**) and atomic-resolution (**c**) STM images of the as-grown pentagonal $PdTe_2$. $V_b$ = 1.0 V, $I_t$ = 0.5 nA for **a**. $V_b$ = -0.5 V, $I_t$ = 0.5 nA for **b** and **c**. Inset of **b**, A schematic image of pentagonal $PdTe_2$. Red solid lines in **b** represent the zigzag chains formed by Te atoms from the topmost layer. The black square in **c** represents the unit cell of pentagonal $PdTe_2$. The direction along zigzag chains is *a* axis. **d**, A STM image of the as-grown pentagonal $PdTe_2$ showing two crystallographic orientations. $V_b$ = -0.5 V, $I_t$ = 0.5 nA. The pink and red solid lines represent the zigzag chains formed by Te atoms on the topmost layer. **e**, A schematic image showing *a* axes of the pentagonal $PdTe_2$ lattices grown along the [$01\bar{2}$] and [$0\overline{12}$] directions of Pd(100) topmost surface, demonstrating the stacking geometry and epitaxial relationship. The dashed black square represents the unit cell of topmost atoms of Pd(100). Black arrows (lower left) indicate the main crystallographic directions of Pd(100). The red and pink squares (solid and dashed) in **d** and **e** represent the unit cells of pentagonal $PdTe_2$ in two orientations. Pink and red arrows in **d** and **e** represent two crystallographic orientations of *a* axes (the direction along zigzag chains) of the as-grown pentagonal

$PdTe_2$. Each orientation has a rotation angle of 26.6° with respect to the unit cell of the topmost atoms of Pd(100). The black dashed lines in **d** and **e** indicate the grain boundary. Note Te adatoms are invisible in **e** because they are covered by bottom Te atoms of $PdTe_2$. **f**, LEED patterns of the as-grown pentagonal $PdTe_2$ measured with a beam energy of 70 eV. Pink and red dashed circles represent the two sets of diffraction patterns. Bright dots without circles are the second order diffraction patterns. LEED patterns from Pd(100) are not observable in this measurement but were measured separately on a bare Pd(100) substrate without changing the crystal orientation (Supplementary Fig. 10).

We carried out further STM measurements and theoretical simulations under various tunneling voltages to explore topographic structures of the pentagonal $PdTe_2$ and electronic effects. Atomically resolved STM images of pentagonal $PdTe_2$ exhibit a strong dependence on tunneling biases (Fig. 4), similar to that of bulk $PdSe_2$[45]. At tunneling biases of 1.0 V and -0.5 V, only features of the zigzag chains formed by topmost Te atoms were resolved (Fig. 4b and 4c). At a large negative tunneling bias of -1.0 V, signals from Pd atoms in the middle layer of $PdTe_2$ became visible (Fig. 4d). This observed bias-dependence could be explained by the calculated total density of states (DOS) (Fig. 4e), where Pd atoms are found to contribute much more strongly to the DOS at more negative energies (relative to the Fermi level). Therefore, the STM image collected at -1.0 V reveals the combination of topographic and electronic effects, while, at positive and small negative energies, the topographic effect is dominant. The simulated STM images at different biases (Fig. 4f-h) are in agreement with the corresponding experimental STM images, confirming the high quality of the as-grown monolayer pentagonal $PdTe_2$.

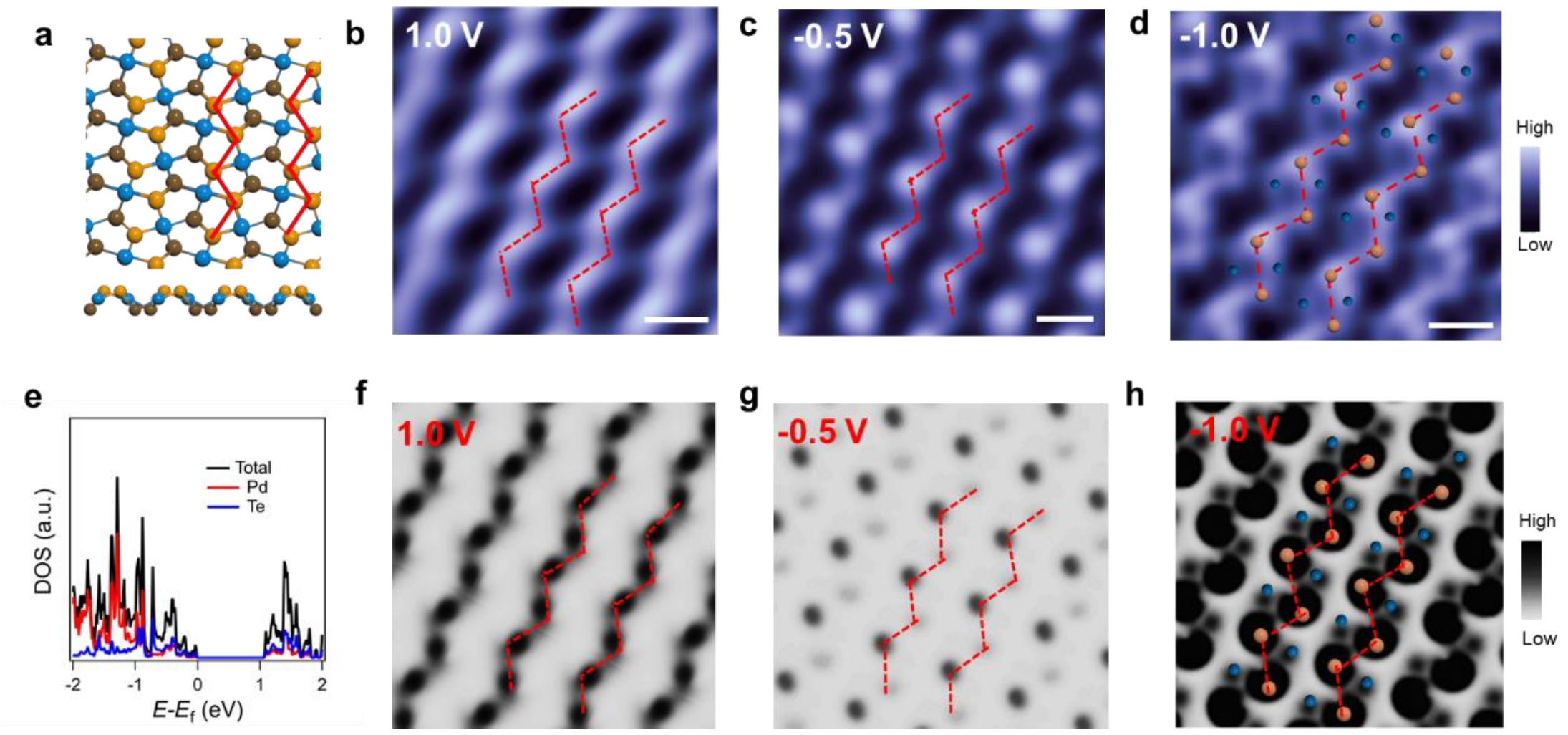

**Fig. 4:** Experimental and simulated STM images at various bias voltages. **a**, Schematic images of monolayer pentagonal $PdTe_2$ in top view (upper) and side view (lower). Orange, blue and brown spheres represent top Te, Pd and bottom Te atoms, respectively. **b-d**, Experimental STM images of the as-grown pentagonal $PdTe_2$ at different biases of 1.0 V, -0.5 V and -1.0 V, respectively. $I_t$ = 0.5 nA. **e**, The calculated DOS for monolayer pentagonal $PdTe_2$. The black curve is the total DOS. The red and blue curves are partial contributions from Pd and Te atoms, respectively. **f-h**, Simulated STM images at different biases of 1.0 V, -0.5 V and -1.0 V, respectively. Red lines (solid and dashed) in all images indicate the zigzag chains formed by Te atoms from the topmost layer. Orange and blue dots in **d** and **h** indicate the Te atoms on the topmost layer and Pd atoms on the middle layer in the $PdTe_2$, respectively. Note the experimental and simulated STM images have opposite color scales.

To further understand the electronic structures, we performed DFT calculations for the electron band structures of monolayer pentagonal $PdTe_2$. The valence band maximum (VBM) of monolayer pentagonal $PdTe_2$ is along the Γ-X path (defined as along the zigzag chain direction in our calculation) while the conduction band minimum (CBM) is along the Γ-Y path (perpendicular to the zigzag chain direction), leading to an indirect bandgap of 1.05 eV (Fig. 5a and Supplementary Fig. 11), which is dramatically different from the narrow indirect bandgap of monolayer hexagonal $PdTe_2$ of 0.33 eV (Supplementary Fig. 12). Valence band structures along three high-symmetry directions, Y-Γ-Y, X-Γ-X and M-Γ-M, were plotted and valence bands dispersed in a "M shape" were observed in all the three directions near the Fermi level (Supplementary Fig. 13). The calculated value of the bandgap and main features of the band structures are comparable with previous calculated results[8,10,11]. To confirm the semiconducting nature and probe the bandgap, we performed STS measurements. The STS curve of the as-grown monolayer pentagonal $PdTe_2$ shows a suppressed conductance region with two onsets of ~ -0.6 V and ~0.6 V, leading to an estimated bandgap of ~1.2 eV (Fig. 5b). This is in reasonable agreement with the calculated 1.05 eV bandgap of monolayer pentagonal $PdTe_2$ (noting DFT calculations are known to underestimate the bandgap[46,47] and were performed on free standing pentagonal $PdTe_2$). Meanwhile the ~1.2 eV bandgap is much bigger than the calculated 0.39 eV bandgap for bilayer and zero bandgap for trilayer pentagonal $PdTe_2$ (Supplementary Fig. 14), further suggesting the as-grown $PdTe_2$ is monolayer. The small nonzero conductance within the bandgap in the STS curve could be attributed to the tunneling to the Pd substrate, as commonly observed in STS curves of monolayer 2D semiconducting materials on metallic substrates[48,49]. We also conducted ARPES measurements to further probe the band structures of monolayer pentagonal $PdTe_2$ (note only valence bands were measured since the conduction bands lie above the Fermi level). ARPES results of the as-grown $PdTe_2$ and clean Pd(100) substrate extracted along a Y-Γ-Y direction (inset of Fig. 5c) are shown in Fig. 5c and 5d, respectively. Since domains of four orientations of pentagonal $PdTe_2$ could be present simultaneously on the surface, band dispersion cutting from the Γ-Y direction of one orientation domain could be superimposed with contributions from domains of other possible orientations, approximately in the Γ-M and Γ-X directions (Supplementary Fig. 15). Given the similarity of the dispersion in the Γ-Y, Γ-M and Γ-X directions (Supplementary Fig. 13), this would merely result in a broadened band of "M shape" valence band. The observed "M-shaped" dispersing band located at ~ -0.65 eV at the Γ point is thus consistent with the calculated valence band structures. This "M shaped" band was also observed along the Γ-X direction (Supplementary Fig. 16). In addition to the broadening by the simultaneous presence of differently-orientated domains, the $PdTe_2$ bands are likely to experience additional broadening and change of dispersion due to the interaction with the substrate. On the other hand, the "M shaped" band clearly retains a two-dimensional character as it does not disperse with different photon energies (Fig. 5e), suggesting the 2D nature of the monolayer pentagonal $PdTe_2$.

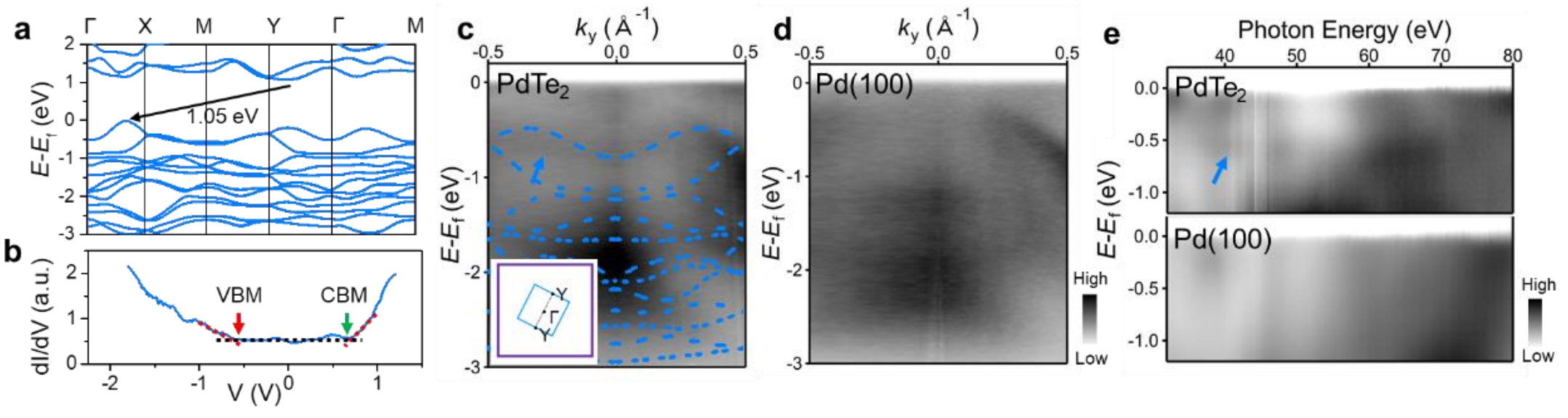

**Fig. 5:** Band structures of monolayer pentagonal $PdTe_2$. **a**, Calculated band structures of monolayer pentagonal $PdTe_2$ with spin-orbital coupling (SOC). **b**, A STS spectrum of monolayer pentagonal $PdTe_2$ on Pd(100). $V_b$ = -100 mV, $I_t$ = 50 pA. The ~1.2 eV bandgap is obtained by measuring the intersections of sharp rises (red dashed lines) and the baseline value of the conductance (black dashed line). Red and green arrows indicate the intersections, corresponding to energies near VBM and CBM, respectively. **c**, Overall ARPES results of monolayer pentagonal $PdTe_2$ extracted along Y-Γ-Y direction (defined for one domain orientation relative to the substrate Brillouin zone as indicated in the inset). The blue arrow points to the “M shape” valence band. The calculated band structures along the Y-Γ-Y direction are superimposed with blue dashed lines. The ARPES result without superimposed calculated band structures is shown in the Supplementary Fig. 17. Inset: Brillouin zones of one pentagonal $PdTe_2$ domain (blue square) and underlying Pd(100) substrate (larger purple square). **d**, ARPES result of clean Pd(100) substrate extracted along the same direction with that in **c**. **e**, Photon energy-dependent spectra at the Γ points for pentagonal $PdTe_2$ (upper panel) and clean Pd(100) (lower panel). The blue arrow points to the non-dispersion pentagonal $PdTe_2$-related band.

In conclusion, we demonstrated the tellurization-based direct synthesis of metastable monolayer pentagonal $PdTe_2$ driven by symmetry epitaxy and provided comprehensive characterizations on this pentagonal 2D material. As confirmed by STM and LEED, the as-grown pentagonal $PdTe_2$ exhibited one-dimensional zigzag atomic chains, with the unit cell matching the $(\sqrt{5}\times\sqrt{5})$R26.6° supercell of the underlying Pd(100) substrate. XPS verified the formation of $PdTe_2$ and suggested its monolayer thickness. Experimentally measured HREELS spectrum probed its lattice vibrations and was consistent with the calculated phonon modes of monolayer pentagonal $PdTe_2$. The experimental STM images of monolayer pentagonal $PdTe_2$ at various bias voltages showed strong bias-dependence, which is also clearly revealed in the simulated STM images. Calculated band structures demonstrated monolayer pentagonal $PdTe_2$ is a semiconductor with an indirect bandgap of 1.05 eV, which agrees with the STS-measured bandgap. The valence band structures were further measured by ARPES, with features consistent with the calculations. The symmetry-driven epitaxial growth provides a viable way for the synthesis of pentagon-based 2D materials, even for metastable phases. Our method will add this unique group of materials to the 2D materials family, so far dominated by hexagon-based ones, and expand it to include a large number of low-symmetry members. This will also bring unprecedented opportunities for 2D materials in applications such as functional electronics based on the in-plane anisotropy, optoelectronics originated from the non-linear light-matter interaction and thermoelectric devices benefited from the low thermal conductivity.

## Acknowledgements

L.L., D.Z., and Y.C. acknowledge partial financial support by the U.S. Department of Energy (Office of Basic Energy Sciences) under Award No. DE-SC0019215 (for synthesis) and Multidisciplinary University Research Initiatives (MURI) Program under Award No. FA9550-20-1-0322 (for characterizations). Y.J. and Y.L. acknowledge support by Natural Science Foundation of Jiangsu Province (BK20200873, BZ2020011) and National Natural Science Foundation of China (Grants No. 22173067). M.B., J.M. and P.H. acknowledge support by VILLUM FONDEN via the Centre of Excellence for Dirac Materials (Grant No. 11744). L.L., R.B., and Y.C. also acknowledge support by VILLUM FONDEN via the Villum Investigator program (Grant No. 25931). Part of the STM measurements (at low temperature) were performed in the Center for Nanophase Materials Sciences (CNMS), which is a US Department of Energy, Office of Science User Facility at Oak Ridge National Laboratory. We express our gratitude to the Center for Nanoscale Materials of Argonne National Laboratory and in particular to Dr. Nathan P Guisinger for the help with the STM tip preparation. We also acknowledge helpful discussions with Espen Drath Bøjesen, Cliff Bugge, Rosa E. Diaz, Yuichi Ikuhara, Lasse Hyldgaard Klausen, Toru Koyama, Akichika Kumatani, Krishna Kanth Neelisetty and the team at Thermo Fisher Scientific, Mitsuhiro Saito, Jim Smith, and Ouyang Yi.

## Author Contributions

L.L. and D.Z. designed the experiments. L.L. and D.Z. performed the experiments and data analysis. Y.J. and Y.L. performed the DFT calculations. M.B., R.B., J.M. and P. H. performed the ARPES measurements and data analysis. S.H and A.L performed the STS measurements and data analysis. Z.L. worked on the synchrotron-based XPS measurements and data analysis. Y.C. coordinated the collaboration and advised the project. L.L. wrote the paper with input from all other co-authors. All authors discussed the results and gave approval of the final version.

## Competing interests

The authors declare no competing interests.

## Methods

**Growth of monolayer pentagonal $PdTe_2$.** Epitaxial growth was conducted in an UHV preparation chamber with a base pressure of $1\times10^{-9}$ mbar. Pd(100) single crystal was used after repeated cycles of $Ar^+$ sputtering (1kV, 10 mA for 10 min at a Ar background pressure of ~$1.0\times10^{-6}$ mbar) and annealing (~700 $^oC$ for 20 min) until clear diffraction patterns were observed by LEED. Tellurium (99.999%, Sigma-Aldrich) was thermally evaporated using a home-built evaporator equipped in the preparation chamber. The substrate was kept at room temperature (RT) during deposition. Te source was kept at ~ 295$^oC$ for 20 min for evaporation. After deposition, annealing was conducted at 500 $^oC$ for 20 min for the $PdTe_2$ growth presented in the main text.

**Characterizations of as-grown $PdTe_2$.** STM, LEED, XPS and HREELS measurements were done at RT (except otherwise noted) in an analysis chamber connected to the preparation chamber with a base pressure of $10^{-11}$ mbar. STM images (Omicron ambient temperature UHV STM) were collected using electrochemically etched W tips at a constant current (topographic) mode (with tips grounded) and were processed using the software WSxM. LEED measurements were conducted with a four-grid optics (Omicron LEED). Lab-based XPS was acquired using a non-monochromatic Mg Kα X-ray radiation ($h\nu$=1253.6 eV) at 150 W. High resolution spectra were recorded at a constant pass energy of 20 eV using the electron energy analyzer-Omicron EAC 125 and the analyzer controller-Omicron EAC 2000. XPS results were analyzed using CasaXPS. The Finite Lorentzian (LF) function and Shirley background function were used for the line shape and background fittings (unless elsewhere noted), respectively. HREELS was performed with ELS5000 from LK technologies. The primary electron beam energy used in this work was 25 eV. The incident and scattering angles were 67°. STS measurements were carried out at 100 K in a separate Omicron variable temperature (VT) STM. Synchrotron-based XPS was collected at the ASTRID2 synchrotron (MatLine) facility at Aarhus University. The MatLine is equipped with an SX-700 monochromator and a SPECS PHOIBOS 150 electron energy analyzer[50]. The analyzer was operated at a pass energy of 20 eV and a curved analyzer slit of 0.8 mm. The beamline monochromator exit slit was set to a width of 30 μm. Photon energies of 410 eV and 80 eV were used for acquiring the Pd 3*d* and Te 4*d* spectra, respectively. ARPES spectra were acquired at the SGM3 beamline[51] at ASTRID2[52], using different photon energies in the range from 20 eV to 130 eV. The data shown in the main text have been acquired using linearly polarized light of 70 eV at 45 K with a combined energy and angle resolution better than 40 meV and 0.2°. The beam spot size on the sample is about 190×90 μm and linearly polarized. The monolayer pentagonal $PdTe_2$ and clean Pd(100) in the synchrotron facility were prepared *in situ* with analogous methods from those presented earlier, with quality and crystallinity cross-checked with a home-built room temperature Aarhus XPS, STM and LEED to confirm the consistency. Fermi surface mapping obtained by ARPES helped to determine the high symmetry crystal axes of $PdTe_2$ relative to the Pd(100) substrate.

**DFT calculations.** All calculations were conducted in the Vienna *ab-initio* Software Package version 5.4.1 in the framework of density functional theory with the basis sets of the projector augmented wave[53-56]. We adopted the revised Perdew-Burke-Ernzerhof (PBE) functional under Pade approximation[57] for geometry optimization in a primitive cell (detailed information is provided in the public repository *https://github.com/Austin6035/DFT-characterization-of-PdTe2-on-Pd*). We set up a model with Te adatoms on the Pd substrate surface (Supplementary Fig. 5b), constituting a buffer layer with Te atoms staying on the hollow sites of Pd(100), between Pd(100) and $PdTe_2$. This model was built because our XPS results (Supplementary Fig. 4b) and a few previous studies suggested the existence of a buffer layer in similar systems (TMDC monolayers epitaxially grow on corresponding single crystals[42,43]). After structural optimization with the model, all calculated results were based on free-standing monolayer pentagonal $PdTe_2$ with the optimized structure. Band structures were calculated with the PBE functional. The kinetic cut-off energy was set to 450 eV and the thresholds of total energy and Feynman-Hellman force during structural optimizations were $10^{-4}$ eV and -0.01 eV/Å, respectively. The Gamma-centered Monkhorst-Pack grids of 4×4×1 were sampled in the Brillouin zones. A vacuum space of 15 Å was added along the z direction to avoid the periodic imaging influence. Phonon dispersions of monolayer pentagonal and hexagonal $PdTe_2$ were calculated with the local density approximation (LDA) functional.